\documentclass[twocolumn]{autart}    

\usepackage{graphicx}
\usepackage{amsmath}
\usepackage{amssymb}
\usepackage{url}
\usepackage{color}
\usepackage{natbib}        
\usepackage{multirow}
\usepackage{hhline}
\usepackage{amsfonts}

\usepackage{subcaption}
\usepackage{bbm}
\usepackage[mathscr]{euscript}

\usepackage[hyperindex,breaklinks]{hyperref}

\newcommand{\f}{\mathscr{F}}
\newcommand{\ld}{\mathscr{L}}

\begin{document}

\begin{frontmatter}

\title{On Topological Conditions for Enabling Transient Control in Leader-follower Networks\thanksref{footnoteinfo}} 

\thanks[footnoteinfo]{This work was supported by the ERC Consolidator Grant LEAFHOUND, the Swedish Research Council (VR) and the Knut och Alice Wallenberg Foundation (KAW).}

\author{Fei Chen}\ead{fchen@kth.se},    
\author{Dimos V. Dimarogonas}\ead{dimos@kth.se}              

\address{Division of Decision and Control Systems, KTH Royal Institute of Technology, SE-100 44, Stockholm, Sweden}  

\begin{keyword}                           
Leader-follower network, Multi-agent systems, Prescribed performance control, Necessary and sufficient condition          
\end{keyword}                             

\begin{abstract}                         
We derive necessary and sufficient conditions for leader-follower multi-agent systems such that we can further apply prescribed performance control to achieve the desired formation while satisfying certain transient constraints. A leader-follower framework is considered in the sense that a group of agents with external inputs are selected as leaders in order to drive the group of followers in a way that the entire system can achieve target formation within certain prescribed performance transient bounds. We first derive necessary conditions on the leader-follower graph topology under which the target formation together with the prescribed performance guarantees can be fulfilled. Afterwards, the derived necessary conditions are extended to necessary and sufficient conditions for leader-follower formation control under transient constraints. Finally, the proposed results are illustrated with simulation examples.
\end{abstract}

\end{frontmatter}

\section{Introduction}\label{sec:introduction}
The problem of decentralized formation control~\citep{fax2004information} of multi-agent systems has been popular due to its wide applications in robotics and manufacturing. Some approaches further consider transient constraints for more complex tasks or certain prescribed system behavior, e.g., connectivity maintenance~\citep{zavlanos2011graph} and collision avoidance~\citep{chang2003collision}. However, most existing approaches apply the designed local control strategies to all agents, which might sometimes be costly or redundant, motivating the need for leader-follower approaches. We thus propose here a systematic way on how to apply transient control in a leader-follower framework, in
which the set of agents with advanced capabilities are selected as leaders to guide the whole agent group to the global equilibrium while fulfilling the transient constraints. 


Control of leader-follower multi-agent systems under transient constraints has been pursued, e.g.,~\cite{katsoukis2016output,dai2019adaptive,he2018leader,wei2018leader}. This standard leader-follower setting typically assumes a single leader which is treated as a reference for the remaining followers and all the follower controllers are designed for the purpose of tracking the leader. It has been extensively studied due to the close relation with the vast literature on multi-agent systems, specifically within a standard leader-follower setting. We instead consider a different leader-follower framework which considers arbitrary number of \emph{leaders} with external inputs in addition to a first-order formation protocol. The remaining agents are \emph{followers} only obeying the first-order formation protocol and have no knowledge or control design freedom for additional objectives, such as the transient behavior pursued here. Therefore, the followers only follow standard cooperative control protocols regardless of the objective of the whole team, while only the leader controllers are designed accordingly to guide the followers such that the target objective is achieved. Recent research in such a leader-follower setting mainly focuses on controllability of leader-follower networks. To name a few,~\cite{tanner2004controllability} derived conditions on the network topology, which ensures that the network can be controlled by  particular members acting as leaders, while an extension of controllability conditions to leader-follower networks with double integrator dynamics was addressed by~\cite{goldin2010controllability}. \cite{egerstedt2012interacting,rahmani2009controllability} investigated necessary conditions for the controllability of the corresponding leader-follower networks using equitable partitions of graphs. ~\cite{sun2017controllability} studied the controllability problem of leader-follower multi-agent systems defined by undirected signed graphs through almost equitable partitions and provided a necessary condition for the controllability of the network. The classes of essentially controllable, completely uncontrollable, and conditionally controllable graphs were discussed by~\cite{aguilar2014graph}. ~\cite{yaziciouglu2016graph,zhang2011controllability,zhang2014upper} derived lower or upper bounds for controllable subspaces of leader-follower networks.
Another category in the leader-follower setup considers leader selection problems to guarantee controllability~\citep{yaziciouglu2013leader} or to maximize a system performance metric~\citep{fitch2016optimal,patterson2016optimal}. Furthermore, ~\cite{pirani2017robustness} studied the robustness of leader–follower consensus dynamics to disturbances and time delays.

As a comparison, we consider transient constraints rather than controllability in the leader-follower setting. Note that controllability is neither sufficient nor necessary to guarantee the fulfillment of the transient constraints, thus the topological conditions to guarantee transient constraints are expected to be different from those for controllability. In addition, the followers are solely indirectly guided through their dynamic couplings with the controlled leaders without any further control and knowledge of the transient control objective in hand, which makes the problem far more challenging than the existing work on transient constraints for (standard leader-follower) multi-agent systems, e.g.,~\cite{katsoukis2016output}. Specifically, a main contribution of the paper in hand is that we consider control of leader-follower multi-agent systems under transient constraints, where the followers do not have any knowledge of the prescribed performance bounds, while only the leader controllers are designed to guide the followers such that the target formation is achieved within the transient constraints. 
Such leader-follower framework finds application in various domains, including multi-vehicle platooning. This involves designing leader vehicle controllers to ensure collision avoidance and maintain connectivity for the entire platoon. In addition, it encompasses multi-robot coordination, involving formation control or even more complex task planning for robot teams under spatiotemporal constraints, where only the controllers of the leader robots are developed. Furthermore, the framework enables collaborative manipulation, involving grasping and transporting objects in a collaborative leader-follower manner while satisfying the transient constraints. All these tasks are assigned to the available leader agents, demonstrating both a distributed and scalable nature. Additionally we investigate the topological conditions on the leader-follower networks such that the target objective can be achieved within the prescribed transient constraints, which offers advantages in various topics, including addressing leader selection problems amidst transient constraints, studying the network's robustness in terms of agent failures, and exploring network reconfigurations. The technical challenges that arise are due to the consideration of transient constraints which are only known to the leaders, and also the combination of non-specific topologies, leader amount and leader positions. Preliminary results of leader-follower formation control with prescribed performance guarantees have been presented by~\cite{chen2020leader,chen2020cdc}, where ~\cite{chen2020cdc} have also briefly discussed necessary conditions on the leader-follower graph topology. In this paper, we present more details on the topological conditions while the necessary conditions ~\citep{chen2020cdc} are further extended to necessary and sufficient conditions on the leader-follower graph topology such that target formation within certain prescribed performance bounds can be achieved. This introduces for the first time a new methodology for leader selection to guarantee target formation while satisfying the transient constraints within the leader-follower framework. 

The contributions and novelty of the paper can be thus summarized as: i) we consider control of leader-follower multi-agent systems under transient constraints, where the followers do not have any knowledge of the prescribed performance bounds; ii) we derive necessary and sufficient conditions on the leader-follower graph topology such that the target formation together with the prescribed transient behavior can be fulfilled for the considered leader-follower multi-agent system.   


\section{Preliminaries and Problem Statement}

\subsection{Graph Theory} 
An undirected graph~\citep{mesbahi2010graph} is defined as $\mathcal{G}=(\mathcal{V},\mathcal{E})$ with the vertices set $\mathcal{V}=\{1,2,\dots,n\}$ and the edges set $\mathcal{E}=\{(i,j)\in \mathcal{V}\times \mathcal{V}\mid j\in \mathcal{N}_i\}$ indexed by $e_1,\dots, e_m$. $m=|\mathcal{E}|$ is the number of edges and $\mathcal{N}_i$ denotes the neighbourhood of agent $i$. For an edge $e_k=(i,j)$, $\mathbf{v}(e_k)=\{i, j\}$ is the set composed of the vertices of $e_k$. The \emph{adjacency matrix} $\mathbb{A}$ of $\mathcal{G}$ is the $n\times n$ symmetric matrix whose elements $a_{ij}$ are given by $a_{ij}=1$, if $(i,j)\in \mathcal{E}$, and $a_{ij}=0$, otherwise. The degree of vertex $i$ is defined as $d(i)={\sum}_{j\in \mathcal{N}_i}a_{ij}$. Then the \emph{degree matrix} is $\Delta=\mbox{diag}(d(1), d(2),\dots,d(n))$. The \emph{graph Laplacian} of $\mathcal{G}$ is $L=\Delta-\mathbb{A}$. 
A \emph{path} is a sequence of edges connecting two distinct vertices. A graph is \emph{connected} if there exists a path between any pair of vertices. By assigning an orientation to each edge of $\mathcal{G}$ the \emph{incidence matrix} $D=D(\mathcal{G})=[d_{ij}]\in \mathbb{R}^{n\times m}$ is defined. The rows of $D$ are indexed by the vertices and the columns are indexed by the edges with $d_{ij}=1$ if the vertex $i$ is the head of the edge $(i,j)$, $d_{ij}=-1$ if the vertex $i$ is the tail of the edge $(i,j)$ and $d_{ij}=0$ otherwise. The \emph{graph Laplacian} of $\mathcal{G}$ is described as $L=DD^T$. $L_e=D^TD$ is the so-called \emph{edge Laplacian}~\citep{zelazo2011edge} and we let $c_{ij}$ denote the elements of $L_e$. 

\subsection{System Description and Proposed Control} 
In this work, we consider a multi-agent system under the communication graph $\mathcal{G}=(\mathcal{V},\mathcal{E})$ with vertices  $\mathcal{V}=\{1,2,\dots,n\}$. Suppose that the first $n_f$ agents are followers while the last $n_l$ agents are  leaders with $\mathcal{V}_\f=\{1,2,\dots,n_f\}$,  $\mathcal{V}_\ld=\{n_f+1,n_f+2,\dots,n_f+n_l\}$ and  $n=n_f+n_l, \mathcal{V}_\f\cap \mathcal{V}_\ld=\emptyset, \mathcal{V}_\f\cup \mathcal{V}_\ld=\mathcal{V}$. For an agent $i$ with $d(i)=1$, we call it an end agent or an end node. Moreover, if it further holds that $i\in \mathcal{V}_\ld$ or $i\in \mathcal{V}_\f$, we call it an end leader and an end follower, respectively.  Here and later on,  the subscript $``\f",``\ld"$ stands for follower and leader vertices set, respectively. We also denote the follower-follower edge set as $\mathcal{E}_{\f\f}=\{e_k\mid 1 \leq k \leq m, k\in \mathbb{Z},e_k=(i,j),i,j\in \mathcal{V}_\f \}$, the leader-leader edge set as $\mathcal{E}_{\ld\ld}=\{e_k\mid 1 \leq k \leq m, k\in \mathbb{Z},e_k=(i,j),i,j\in \mathcal{V}_\ld \}$ and the leader-follower edge set as $\mathcal{E}_{\ld\f}=\mathcal{E}\setminus(\mathcal{E}_{\f\f}\cup \mathcal{E}_{\ld\ld})$. In a similar manner, we can define $\mathcal{V}_\f^{\text{sup}}\subseteq \mathcal{V}_\f, \mathcal{V}_\ld^{\text{sup}}\subseteq \mathcal{V}_\ld, \mathcal{E}_{\f\f}^{\text{sup}}\subseteq \mathcal{E}_{\f\f}, \mathcal{E}_{\ld\ld}^{\text{sup}}\subseteq \mathcal{E}_{\ld\ld}, \mathcal{E}_{\ld\f}^{\text{sup}}\subseteq \mathcal{E}_{\ld\f}$ with respect to the subgraphs $\mathcal{G}^{\text{sup}}=(\mathcal{V}^{\text{sup}},\mathcal{E}^{\text{sup}})$ of $\mathcal{G}$ over different  superscripts, e.g., we use ${\text{sup}}\in \{f, p, \star, \prime, \cdots \}$ later on.

Let $p_i\in \mathbb{R}$ be the position of agent $i$, where we only consider the one dimensional case, without loss of generality. Specifically, the results can be extended to higher dimensions with Kronecker product. The target relative position-based formation is described as follows: 
\begin{equation}\label{eq:formation}
\mathcal{F}:=\{p\mid p_i-p_j=p^{des}_{ij},(i,j)\in\mathcal{E}\},    
\end{equation}
where $p^{des}_{ij}:=p^{des}_i-p^{des}_j, (i,j)\in \mathcal{E}$ is the desired relative position between agent $i$ and agent $j$, which is constant and denoted as the difference between the absolute desired positions $p^{des}_i, p^{des}_j\in \mathbb{R}$. The state evolution of each agent $i\in \mathcal{V}$ is governed by the following dynamics:
\begin{equation}\label{eq:ldynamic}
\dot{p}_i=-\sum\limits_{j\in \mathcal{N}_i}(p_i-p_j-p^{des}_{ij})+b_iu_i,
\end{equation}
with $b_i=1$ if $i\in \mathcal{V}_\ld$, and $b_i=0$ if $i\in \mathcal{V}_\f$. This means that followers are governed by the first-order formation protocol, while leaders are governed by the first-order formation protocol with an assigned external input. By stacking \eqref{eq:ldynamic}, the dynamics of the leader-follower multi-agent system are rewritten as:
\begin{equation}\label{eq:nodedynamic}
  \Sigma:   \dot{p}=-L(p-p^{des})+Bu,
\end{equation}
where $p=[p_1,\dots,p_n]^T, p^{des}=[p^{des}_1,\dots,p^{des}_n]^T\in \mathbb{R}^n$, $u=[u_{n_f+1},\dots,u_{{n_f+n_l}}]^T\in \mathbb{R}^{n_l}$ and $B=\left[
\begin{smallmatrix} 0_{n_f\times n_l}\\ I_{n_l}
\end{smallmatrix}\right].$ Denote $\bar{p}=[\bar{p}_1,\dots,\bar{p}_m]^T$, $ \bar{p}^{des}=[\bar{p}^{des}_1,\dots,\bar{p}^{des}_m]^T\in \mathbb{R}^m$ as the respective stack vector of relative positions and target relative positions between the pair of communication agents for the edge $(i,j)=e_k\in \mathcal{E}$, where $\bar{p}_k\triangleq p_{ij}=p_i-p_j,  \bar{p}^{des}_k\triangleq p^{des}_{ij}=p^{des}_i-p^{des}_j, k=1,2,\dots, m$. Accordingly, $\bar{x}=\bar{p}-\bar{p}^{des}=[\bar{x}_1,\dots,\bar{x}_m]^T$ is denoted as the shifted relative position vector. The incidence matrix $D$ can be decomposed by the rows into the first $n_f$ and the remaining last $n_l$ rows, i.e.,  $D=\begin{bmatrix}
D_\f^T&D_\ld^T
\end{bmatrix}^T$. Multiplying with $D^T$ on both sides of \eqref{eq:nodedynamic}, the dynamics \eqref{eq:nodedynamic} can be reorganized into the edge space as 
\begin{equation}\label{eq:DS_edge}
\Sigma_{e}: \dot {\bar{x}}=-L_e\bar{x}+D_\ld^Tu,
\end{equation}
The aim of prescribed performance control (PPC)~\citep{bechlioulis2008robust} is to prescribe the evolution of the relative position $\bar{p}_i(t)$ within some predefined region described as
\begin{equation}\label{eq:pfp}
\bar{p}^{des}_i-\rho_{\bar{x}_i}(t)<\bar{p}_i(t)<\bar{p}^{des}_i+\rho_{\bar{x}_i}(t),
\end{equation}
or equivalently, to prescribe $\bar{x}_i(t)$ within
$
-\rho_{\bar{x}_i}(t)<\bar{x}_i(t)<\rho_{\bar{x}_i}(t).
$ Here $\rho_{\bar{x}_i}(t):\mathbb{R}_+\rightarrow \mathbb{R}_+\setminus \{0\}, i=1,2,\dots,m$ are positive, smooth and strictly decreasing performance functions and one example choice is $
\rho_{\bar{x}_i}(t)=(\rho_{\bar{x}_{i0}}-\rho_{\bar{x}_{i\infty}})e^{-l_{\bar{x}_i}t}+\rho_{\bar{x}_{i\infty}}
$
with $\rho_{\bar{x}_{i0}},\rho_{\bar{x}_{i\infty}}$ and $l_{\bar{x}_i}$ being positive parameters. We further define the modulated error as $\hat{\bar{x}}_{i}(t)=\frac{\bar{x}_i(t)}{\rho_{\bar{x}_i}(t)}$ and the corresponding prescribed performance region as
$
\mathcal{D}_{\bar{x}_i}\triangleq \{\hat{\bar{x}}_{i}:\hat{\bar{x}}_{i}\in (-1,1)\}.
$
Then the modulated error is transformed through a function $T_{\bar{x}_i}$ that defines the smooth and strictly increasing mapping $T_{\bar{x}_i}: \mathcal{D}_{\bar{x}_i}\rightarrow \mathbb{R}$, $T_{\bar{x}_i}(0)=0$. Here, one example choice is 
  $  T_{\bar{x}_i}(\hat{\bar{x}}_{i})=\ln \left(\frac{1+\hat{\bar{x}}_{i}}{1-\hat{\bar{x}}_{i}}\right).$ The transformed error is then defined as 
$
\varepsilon_{\bar{x}_i}(\hat{\bar{x}}_{i})=T_{\bar{x}_i}(\hat{\bar{x}}_{i}). 
$
Differentiating $\varepsilon_{\bar{x}_i}(\hat{\bar{x}}_{i})$ with respect to time, we derive
$\dot{\varepsilon}_{\bar{x}_i}(\hat{\bar{x}}_{i})=\mathcal{J}_{T_{\bar{x}_i}}(\hat{\bar{x}}_{i},t)(\dot{\bar{x}}_{i}+\alpha_{\bar{x}_i}(t)\bar{x}_i),$
where
$\mathcal{J}_{T_{\bar{x}_i}}(\hat{\bar{x}}_{i},t)\triangleq \frac{\partial T_{\bar{x}_i}(\hat{\bar{x}}_{i})}{\partial \hat{\bar{x}}_{i}}\frac{1}{\rho_{\bar{x}_i}(t)}$ and
$\alpha_{\bar{x}_i}(t)\triangleq -\frac{\dot{\rho}_{\bar{x}_i}(t)}{\rho_{\bar{x}_i}(t)}$
are the normalized Jacobian of the transformed function $T_{\bar{x}_i}$ and the normalized derivative of the performance function, respectively. The basic idea of PPC is to verify that the transformed error $\varepsilon_{\bar{x}_i}(\hat{\bar{x}}_{i})$ is bounded, which in turn results in the satisfaction of \eqref{eq:pfp}~\citep{bechlioulis2008robust}. For the edge dynamics \eqref{eq:DS_edge}, the following PPC control strategy is proposed by~\cite{chen2020leader,chen2020cdc}:
\begin{equation}\label{eq:control}
u_j=-\sum\limits_{i\in \Phi_j}g_{\bar{x}_i}\mathcal{J}_{T_{\bar{x}_i}}(\hat{\bar{x}}_i,t)\varepsilon_{\bar{x}_i}(\hat{\bar{x}}_i), \hspace{5mm} j\in \mathcal{V}_\ld,
\end{equation} 
where $\Phi_j=\{i|(j,k)=i,k\in\mathcal{N}_j\}$, i.e., the set of all the edges that include agent $j\in\mathcal{V}_\ld$ as a node, and  $g_{\bar{x}_i}$ is a positive scalar gain to be appropriately tuned. It is concluded by~\cite{chen2020leader,chen2020cdc} that the leader-follower multi-agent system \eqref{eq:nodedynamic} can achieve the target formation $\mathcal{F}$ as in \eqref{eq:formation} while satisfying \eqref{eq:pfp} by applying \eqref{eq:control} under some assumptions on the graph topology, which will be further discussed in this paper. We refer the readers to~\cite{chen2020leader,chen2020cdc} for the detailed convergence analysis. 

\subsection{Problem Statement}
In this work, we focus on investigating the leader-follower graph topology and the following problem is formulated.
\begin{prob}
Derive the necessary and sufficient conditions on the leader-follower communication graph $\mathcal{G}=(\mathcal{V},\mathcal{E})$ such that the leader-follower multi-agent system \eqref{eq:nodedynamic} achieves the target formation $\mathcal{F}$ as in \eqref{eq:formation} while satisfying \eqref{eq:pfp} by applying the proposed control strategy \eqref{eq:control}.
\end{prob}

\section{Necessary Conditions on Graph Topology}
In this section, we derive necessary conditions on the graph topology for both tree graphs and general graphs with cycles such that under these conditions we can design the leaders to achieve the target formation with prescribed performance guarantees.  

We first discuss the tree graphs and then the results for general graphs with cycles are built based on the results of tree graphs. We first define a \emph{leaderless graph} $\mathcal{G}^f=(\mathcal{V}^f,\mathcal{E}^f)$ with only followers, i.e., $\mathcal{V}_\f^f=\mathcal{V}^f$, $\mathcal{V}_\ld^f=\emptyset$ and $\mathcal{E}_{\f\f}^f=\mathcal{E}^f$. The insight here is that we would like to analyze how the leader-follower multi-agent system described by the graph $\mathcal{G}=(\mathcal{V},\mathcal{E})$ behaves when it contains $\mathcal{G}^f$ as an induced subgraph. The definition of subgraph and induced subgraph is given as follows:
\begin{defn}\label{def:sub}
(Subgraph and induced subgraph.) A graph $\mathcal{G}^{\prime}=(\mathcal{V}^{\prime},\mathcal{E}^{\prime})$ is a subgraph of the graph $\mathcal{G}=(\mathcal{V},\mathcal{E})$ if $\mathcal{V}^{\prime}\subseteq \mathcal{V}$ and $\mathcal{E}^{\prime}\subseteq \mathcal{E}$;
for any $i\in \mathcal{V}_\f^{\prime}$ we have $i\in \mathcal{V}_\f$ and for any $i\in \mathcal{V}_\ld^{\prime}$ we have $i\in \mathcal{V}_\ld$. A subgraph $\mathcal{G}^{\prime}$ of $\mathcal{G}$ is an induced subgraph of $\mathcal{G}$, denoted as $\mathcal{G}^{\prime} \subseteq \mathcal{G}$ if it further holds that for any edge $(i,j)\in \mathcal{E}, i,j \in \mathcal{V}^\prime$, we have $(i,j)\in \mathcal{E}^\prime$.
\end{defn}
\begin{defn} \label{def:path}
(Path subgraph.) A path $p$ of the graph $\mathcal{G}=(\mathcal{V},\mathcal{E})$ is called a path subgraph $p=(\mathcal{V}^{p},\mathcal{E}^{p})$ of $\mathcal{G}$ such that 
     $\mathcal{V}^{p}\subseteq \mathcal{V}$ and $\mathcal{E}^{p}\subseteq \mathcal{E}$,
     for any $i\in \mathcal{V}_\f^{p}$ we have $i\in \mathcal{V}_\f$ and for any $i\in \mathcal{V}_\ld^{p}$ we have $i\in \mathcal{V}_\ld$;
\end{defn}
From now on, we denote $\bar{x}$ as the edge state of $\mathcal{G}^f$ and $L_e$ as the edge Laplacian of $\mathcal{G}^f$. We know that the edge dynamics of $\mathcal{G}^f$ are simply described as $\dot{\bar{x}}=-L_e\bar{x}$ since the leader set $\mathcal{V}^f_\ld$ is empty. Denote each column (corresponding to an edge) of the incidence matrix of $\mathcal{G}^f$ by the vector $e_i$. Then $(L_e)_{ij}=e_i^Te_j=c_{ij}=2$ if $i=j$; $c_{ij}=0$ if $e_i,e_j$ share no nodes; $c_{ij}=1$ if $e_i,e_j$ share a single node and have the same direction with respect to the sharing node; $c_{ij}=-1$ if $e_i,e_j$ share a single node but have different direction with respect to the sharing node~\citep{zelazo2011edge}. Based on these simple rules, we can derive the dynamics in the edge space. Example \ref{exmp:1} elucidates such a derivation. We define the neighbors of edge $e_i$ as $\mathcal{N}(e_i):=\{e_j\mid |e_i^Te_j|=1\}.$
\begin{exmp}\label{exmp:1}
We consider a leaderless tree graph $\mathcal{G}^f$ as in Fig. \ref{exmp_edge} where we assign arbitrary directions for all the edges in order to derive the corresponding incidence matrix $D$ and the edge Laplacian $L_e$ (note that we still focus on undirected graphs). Then the edge dynamics of the system described by $\mathcal{G}^f$ are $\dot{\bar{x}}=-L_e\bar{x}=-D^TD\bar{x}$ with $\bar{x}=\begin{bmatrix}
    \bar{x}_1&\bar{x}_2&\bar{x}_3&\bar{x}_4
    \end{bmatrix}^T$ and $D, L_e$ matrices as follows:  

\begin{equation}\nonumber
    D=\left[\begin{smallmatrix}
    -1&1&1&1\\1&0&0&0\\0&-1&0&0\\0&0&-1&0\\0&0&0&-1
    \end{smallmatrix}\right], L_e=\left[\begin{smallmatrix}
    2&-1&-1&-1\\-1&2&1&1\\-1&1&2&1\\-1&1&1&2
    \end{smallmatrix}\right].
\end{equation}
So the dynamics of the edge $e_1$ are $\dot{\bar{x}}_1=-2\bar{x}_1+\bar{x}_2+\bar{x}_3+\bar{x}_4$ according to first row of $L_e$. Based on above discussions, we can also derive $\dot{\bar{x}}_1$ by directly checking the graph only. In Fig. \ref{exmp_edge}, edge $e_1$ has three neighboring edges $e_2, e_3, e_4$ since they share the node indexed by $1$. Then we denote  $\dot{\bar{x}}_1=a\bar{x}_1+b\bar{x}_2+c\bar{x}_3+d\bar{x}_4$ with $\begin{bmatrix}
    -a&-b&-c&-d
    \end{bmatrix}$ being the first row of $L_e$. We know that $a=-2$ since the diagonal entry of $L_e$ is always $2$. $b=c=d=1$ since edge $e_1$ shares node $1$ with edges $e_2, e_3, e_4$ butt have different directions with respect to node $1$ ($e_1$ points outward node $1$, $e_2, e_3, e_4$ all point inward node $1$, and we assign $-1$ to the corresponding entries of $L_e$). 

\begin{figure}[!h]
\centering
\includegraphics[width=0.3\columnwidth]{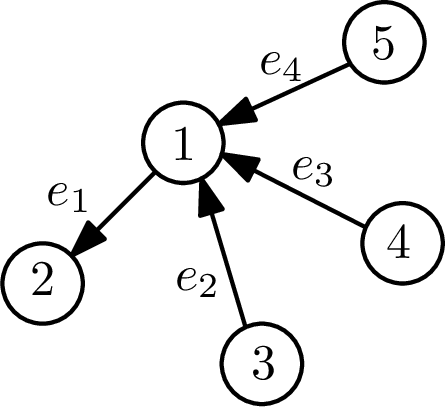}
\caption{Example of dynamics for a tree graph via edge Laplacian.}
\label{exmp_edge}
\end{figure}
\end{exmp}

Then the following lemma is proposed for tree graphs and acts as a basis for the later extended results for general graphs with cycles.  
\begin{lem}\label{lem:1}
Consider the leader-follower multi-agent system $\Sigma$ described by the tree graph $\mathcal{G}=(\mathcal{V},\mathcal{E})$. A necessary condition on $\mathcal{G}$ under which we can design the leaders using \eqref{eq:control} to achieve the target formation $\mathcal{F}$ as in \eqref{eq:formation} while satisfying \eqref{eq:pfp} is that every leaderless graph $\mathcal{G}^f=(\mathcal{V}^f,\mathcal{E}^f)$, such that $\mathcal{G}^f \subseteq \mathcal{G} $, should satisfy
\begin{equation}\label{eq:ne_tree}
    |\mathcal{N}(e_i)|\leq 2, \hspace{3mm}\forall e_i\in \mathcal{E}^f.
\end{equation}
\end{lem}
\begin{pf}
The proof uses contradiction based on the entries of the edge Laplacian $L_e$ of $\mathcal{G}^f$. Suppose $\mathcal{G}^f$ is a leaderless induced subgraph of $\mathcal{G}$ and there exists $ e_i\in \mathcal{E}^f$ satisfying $|\mathcal{N}(e_i)|\geq 3$. Without loss of generality, let us assume that $e_1\in \mathcal{E}^f$ satisfies $\mathcal{N}(e_1)=\{e_2,e_3,e_4\}$, thus $|\mathcal{N}(e_i)|= 3$. Suppose that $e_2,e_3,e_4$ all share a single node with $e_1$ but with different directions as shown in Fig. \ref{exmp_edge}. This can be assumed without loss of generality since we can assign arbitrary directions to the edges. Then the state evolution of $e_1$ is derived as $\dot{\bar{x}}_1=-2\bar{x}_1+\bar{x}_2+\bar{x}_3+\bar{x}_4$ according to Example \ref{exmp_edge}. We can see that when all $\bar{x}_i, i=1,2,3,4$ are initialised arbitrarily close to the prescribed performance bound $\rho_{\bar{x}_{i0}}=\rho_0, i=1,2,3,4$, then $\dot{\bar{x}}_1=\rho_0>0$, thus $\bar{x}_1$ will continue evolving to violate the performance bound. This leads to a contradiction since no matter how we design the leaders in $\mathcal{G}$, $\bar{x}_1$ will always increase to violate the bound. Hence, we can conclude that $\mathcal{G}$ should not contain a leaderless induced subgraph $\mathcal{G}^f=(\mathcal{V}^f,\mathcal{E}^f)$ with $\mathcal{V}_F^f=\mathcal{V}^f$ such that there exists $ e_i\in \mathcal{E}^f$ satisfying $|\mathcal{N}(e_i)|\geq 3$. Or in other words, every leaderless graph $\mathcal{G}^f=(\mathcal{V}^f,\mathcal{E}^f)$, such that $\mathcal{G}^f \subseteq \mathcal{G} $, should satisfy $|\mathcal{N}(e_i)|\leq 2, \forall e_i\in \mathcal{E}^f.$\qed
\end{pf}
Next, based on Lemma 1, we derive a necessary condition for graphs with cycles. We denote here $\mathcal{E}(\mathcal{C})$ as the edge set of the cycle $\mathcal{C}$ with cardinality $|\mathcal{E}(\mathcal{C})|$.  We then perform the following graph decomposition which we call \emph{complete decomposition}. 
\begin{defn}\label{def:cd_edge}
(Complete decomposition.) A graph $\mathcal{G}=(\mathcal{V},\mathcal{E})$ is decomposed with respect to the edge $e_i\in \mathcal{E}$ as $\mathcal{G}:=\cup_{\mathcal{C}_{e_i}\in \mathcal{X}_i}\mathcal{C}_{e_i}\cup \mathcal{P}_i$, where $\mathcal{X}_i:=\{\mathcal{C}_{e_i}\mid e_i\in \mathcal{C}_{e_i}\}$ is the cycle set composed of all the cycles $\mathcal{C}_{e_i}$ in $\mathcal{G}$ that contain $e_i$ as an edge, and satisfy:
\begin{itemize}
\item for every pair $\mathcal{C}_{e_i}^a, \mathcal{C}_{e_i}^b\in \mathcal{X}_i$, $(\mathcal{N}(e_i)\cap\mathcal{C}_{e_i}^a)\cap (\mathcal{N}(e_i)\cap\mathcal{C}_{e_i}^b)=\emptyset$.
\item for every $\mathcal{C}_{e_i} \in \mathcal{X}_i$, there does not exist a cycle $\mathcal{C}$ of $\mathcal{G}$ such that $e_i\in \mathcal{C}, (\mathcal{N}(e_i)\cap\mathcal{C}_{e_i})\cap (\mathcal{N}(e_i)\cap\mathcal{C})\neq \emptyset$, and $|\mathcal{E}(\mathcal{C})|<|\mathcal{E}(\mathcal{C}_{e_i})|$,
\end{itemize}
and where $\mathcal{P}_i:=\{e_k\mid e_k\notin \mathcal{C}_{e_i}, \mathcal{C}_{e_i} \in \mathcal{X}_i\}$ is the set of the edges that do not belong to any cycle in $\mathcal{X}_i$. Then, we call this decomposition a complete decomposition of $\mathcal{G}$ with respect to the edge $e_i\in \mathcal{E}$.  
\end{defn}
Next, we elucidate the complete decomposition with the following example.
\begin{exmp}
Consider two leaderless graphs as shown in Fig.\ref{exmp1}. In the left figure, the complete decomposition with respect to the edge $e_1$ is $\mathcal{G}:=\cup_{\mathcal{C}_{e_1}\in \mathcal{X}_1}\mathcal{C}_{e_1}\cup \mathcal{P}_1$, where the cycle set $\mathcal{X}_1$ includes only the cycle on the top, i.e., the cycle composed by the edges $e_1,e_2,e_3$, and  $\mathcal{P}_1=\{e_4,e_5,e_6\}$. The complete decomposition with respect to the edge $e_3$ is $\mathcal{G}:=\cup_{\mathcal{C}_{e_3}\in \mathcal{X}_3}\mathcal{C}_{e_3}\cup \mathcal{P}_3$, where the cycle set $\mathcal{X}_3$ includes the cycle composed by the edges $e_1,e_2,e_3$ and the cycle composed by the edges $e_3,e_4,e_5,e_6$, and thus $\mathcal{P}_3=\emptyset$. The second condition of the cycle set in Definition 3 implies that we do not consider the large cycle $\mathcal{C}^\prime$ composed of the edges $e_1,e_2,e_6,e_5,e_4$ as in the cycle set $\mathcal{X}_1$ or $\mathcal{X}_3$. For instance, consider $e_1$ again, and let us assume $\mathcal{C}^\prime\in \mathcal{X}_1$. Then there exists a cycle $\mathcal{C}$ which is composed by $e_1,e_2,e_3$ such that $e_1\in \mathcal{C}, (\mathcal{N}(e_1)\cap\mathcal{C}^\prime)\cap (\mathcal{N}(e_1)\cap\mathcal{C})=\{e_2\}\neq \emptyset$ since $\mathcal{N}(e_1)=\{e_2,e_3,e_4\}$. In addition $|\mathcal{E}(\mathcal{C})|=3<|\mathcal{E}(\mathcal{C}^\prime)|=5$, which contradicts the second condition of Definition 2. Therefore, we can conclude that  $\mathcal{C}^\prime\notin \mathcal{X}_1$. 
In the right figure, the complete decomposition with respect to the edge $e_4$ is $\mathcal{G}:=\cup_{\mathcal{C}_{e_4}\in \mathcal{X}_4}\mathcal{C}_{e_4}\cup \mathcal{P}_4$, where the cycle set $\mathcal{X}_4$ includes the cycle composed by the edges $e_1,e_4,e_5$ and the cycle composed by the edges $e_3,e_4,e_6$; $\mathcal{P}_4=\{e_2,e_7\}$ and thus $\mathcal{N}(e_4)\cap \mathcal{P}_4=\{e_7\}$.
\begin{figure}[!h]
\centering
\includegraphics[width=0.7\columnwidth]{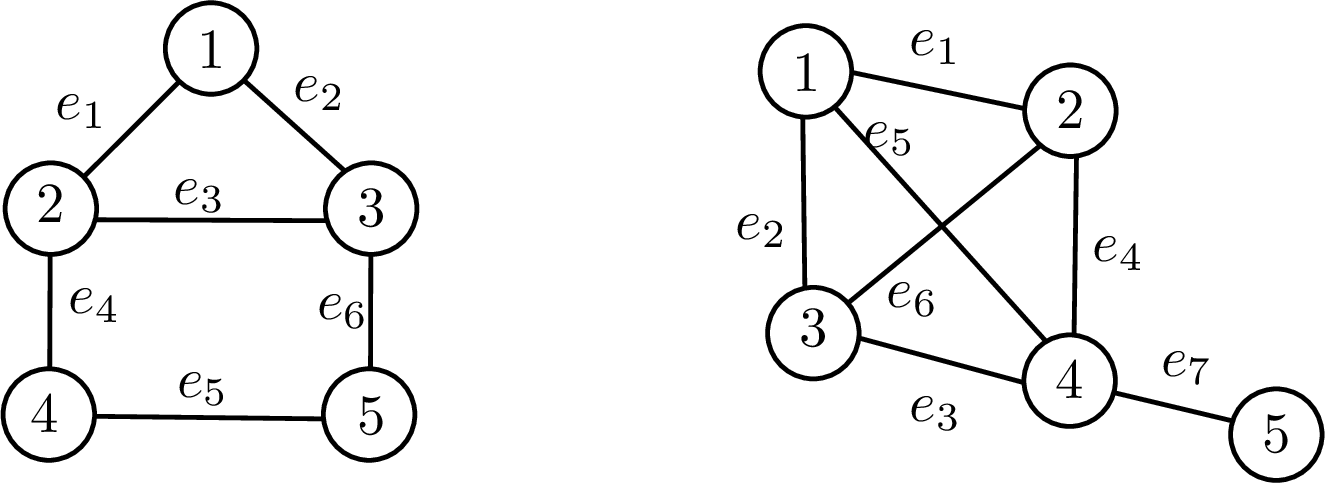}
\caption{Complete decomposition examples}
\label{exmp1}
\end{figure}
\end{exmp}
The complete decomposition decomposes a large scale graph into certain cycles together with the remaining edges that do not belong to any cycle. We can then derive the necessary condition for $\mathcal{G}$ based on the decomposed cycle set $\mathcal{X}_i$ with respect to the edge $e_i\in \mathcal{E}$ and the remaining edges in $\mathcal{P}_i$. This is done based on analysing the convergence of each edge, which is affected by its neighboring edges. Thus the insights here of the complete decomposition are: for the first requirement of Definition \ref{def:cd_edge}, we avoid considering the neighboring edges repeatedly; for the second requirement of Definition \ref{def:cd_edge}, we consider the smallest cycle that contains the edge in hand.   
The following theorem proposes a necessary condition on general graphs with cycles. 
\begin{thm}\label{thm:1}
Consider the leader-follower multi-agent system $\Sigma$ described by the graph $\mathcal{G}=(\mathcal{V},\mathcal{E})$. A necessary condition on $\mathcal{G}$ under which we can design the leaders using \eqref{eq:control} to achieve the target formation $\mathcal{F}$ as in \eqref{eq:formation} while satisfying \eqref{eq:pfp} is that every leaderless graph $\mathcal{G}^f=(\mathcal{V}^f,\mathcal{E}^f)$, such that $\mathcal{G}^f \subseteq \mathcal{G} $, should satisfy
\begin{equation}\label{eq:ne_cycle}
    \sum\limits_{\mathcal{C}_{e_i}\in \mathcal{X}_i}\left\{\min (|\mathcal{E}(\mathcal{C}_{e_i})|-4,2)\right\}+|E_i|\leq 2, \hspace{3mm}\forall e_i\in \mathcal{E}^f.
\end{equation}
where $E_i=\{e_k\mid e_k \in \mathcal{N}(e_i)\cap \mathcal{P}_i \}$, and $\mathcal{X}_i$ is the cycle set obtained via the complete decomposition of $\mathcal{G}^f$ with respect to $e_i$ as in Definition \ref{def:cd_edge}.
\end{thm}
\begin{pf}
The proof is based on the discussion of the decomposed cycles $\mathcal{C}_{e_i}\in \mathcal{X}_i$ and the remaining edges in $\mathcal{P}_i$. We can resort to Lemma \ref{lem:1} to deal with $\mathcal{P}_i$. Here, we first discuss the result for one cycle, e.g., $\mathcal{C}_{e_1}^a$ with respect to the edge $e_1$ which has $m$ edges $\mathcal{E}(\mathcal{C}_{e_1}^a)=\{e_1,e_2,\dots,e_m\}$ and $m\geq 3$ (need at least 3 edges to form a cycle). We will check how the number of edges of the cycle affects convergence. When $m=3$, the state evolution of an arbitrary edge $e_1$ is $\dot{\bar{x}}_1=-2\bar{x}_1+\bar{x}_2+\bar{x}_3$, and since $e_i,i=1,2,3$ form a cycle, we have that $\bar{x}_1+\bar{x}_2+\bar{x}_3=0$. Hence $\bar{x}_2+\bar{x}_3=-\bar{x}_1$ and  $\dot{\bar{x}}_1=-2\bar{x}_1-\bar{x}_1=-3\bar{x}_1$. This means that the cycle that forms a triangle will show a higher convergence rate of $-3$ for the edge dynamics. When $m=4$, the state evolution of an arbitrary edge $e_1$ is $\dot{\bar{x}}_1=-2\bar{x}_1+\bar{x}_2+\bar{x}_4$, and since $e_i,i=1,2,3,4$ form a cycle, we have that $\sum_{i=1}^4\bar{x}_i=0$. Hence $\bar{x}_2+\bar{x}_4=-\bar{x}_1-\bar{x}_3$ and  $\dot{\bar{x}}_1=-2\bar{x}_1-\bar{x}_1-\bar{x}_3$. Consider the worst case, when $\bar{x}_1$ is arbitrarily close to the prescribed performance bound and $\bar{x}_3=-\bar{x}_1$, we then have $\dot{\bar{x}}_1=-2\bar{x}_1-\bar{x}_1+\bar{x}_1=-2\bar{x}_1$. This means that the cycle still shows decay rate of $-2$ for the edge dynamics. When $m=5$, the state evolution of an arbitrary edge $e_1$ is $\dot{\bar{x}}_1=-2\bar{x}_1+\bar{x}_2+\bar{x}_5$, and since $e_i,i=1,2,3,4,5$ form a cycle, we have that $\sum_{i=1}^5\bar{x}_i=0$. Hence $\bar{x}_2+\bar{x}_5=-\bar{x}_1-\bar{x}_3-\bar{x}_4$ and  $\dot{\bar{x}}_1=-2\bar{x}_1-\bar{x}_1-\bar{x}_3-\bar{x}_4$. Consider the worst case, when $\bar{x}_1$ is arbitrarily close to the performance bound and $\bar{x}_3=\bar{x}_4=-\bar{x}_1$, we then have $\dot{\bar{x}}_1=-2\bar{x}_1-\bar{x}_1+2\bar{x}_1=-\bar{x}_1$. This means that the cycle with $5$ edges still shows decay rate of $-1$ for the edge dynamics. This means that we still have the freedom to add one more edge that shares a node with $e_1$. For $m\geq6$, the state evolution of an arbitrary edge $e_1$ is $\dot{\bar{x}}_1=-2\bar{x}_1+\bar{x}_2+\bar{x}_m$, and since $e_i,i=1,2,\dots,m$ forms a cycle, we have that $\sum_{i=1}^m\bar{x}_i=0$. Then $\bar{x}_2+\bar{x}_m=-\bar{x}_1-\sum_{i=3}^{m-1}\bar{x}_i$ and $\dot{\bar{x}}_1=-2\bar{x}_1-\bar{x}_1-\sum_{i=3}^{m-1}\bar{x}_i$. $-\sum_{i=3}^{m-1}\bar{x}_i$ cannot be greater than $3\bar{x}_1$ when $\bar{x}_1$ is arbitrarily close to the performance bound since  $e_i,i=1,2,\dots,m$ need to form a cycle satisfying $\sum_{i=1}^m\bar{x}_i=0$. This means that for the cycle with $m\geq6$, in the worst case we have $\dot{\bar{x}}_1=0$, thus in the worst case  $\bar{x}_1$ will never evolve again to violate the performance bound. Till here, we can summarize that for a single cycle $\mathcal{C}_{e_1}^a$, in the worst case the decay rate of the edge dynamics is $-2+\min(|\mathcal{E}(\mathcal{C}_{e_1}^a)|-4,2)$. Then, based on the result for a single cycle, we build the result in the case that $e_i$ belongs to more than one cycles, i.e., $e_i$ is an edge of $\mathcal{C}_{e_i}\in \mathcal{X}_i$. Then each cycle  of $\mathcal{C}_{e_i}\in \mathcal{X}_i$ will contribute a decay rate of $\min(|\mathcal{E}(\mathcal{C}_{e_i})|-4,2)$ for $e_i$. Since the cycles $\mathcal{C}_{e_i}\in \mathcal{X}_i$ will not affect each other in contributing to the decay rate of $e_i$ since they are completely decomposed with respect to $e_i$, then the total decay rate of $e_i$ that belongs to more than one cycle is $-2+\sum\limits_{\mathcal{C}_{e_i}\in \mathcal{X}_i}\left\{\min (|\mathcal{E}(\mathcal{C}_{e_i})|-4,2)\right\}$, where $-2$  corresponds to the diagonal entry of the edge Laplacian $-L_e$. Finally, the remaining edges that affect the convergence of $e_i$ are the edges that share a node with $e_i$ but are not an edge of any $\mathcal{C}_{e_i}\in \mathcal{X}_i$, i.e., $E_i=\{e_k\mid e_k \in \mathcal{N}(e_i)\cap \mathcal{P}_i \}$. In the worst case, each of these edges will contribute a decay rate of $1$ to edge $e_i$. Hence, the total decay rate of $e_i$ in the worst case is $-2+\sum\limits_{\mathcal{C}_{e_i}\in \mathcal{X}_i}\left\{\min (|\mathcal{E}(\mathcal{C}_{e_i})|-4,2)\right\}+|E_i|$ and should satisfy that $-2+\sum\limits_{\mathcal{C}_{e_i}\in \mathcal{X}_i}\left\{\min (|\mathcal{E}(\mathcal{C}_{e_i})|-4,2)\right\}+|E_i|\leq 0$, i.e., exactly the inequality in \eqref{eq:ne_cycle}. This means that in the worst case when the term $\sum\limits_{\mathcal{C}_{e_i}\in \mathcal{X}_i}\left\{\min (|\mathcal{E}(\mathcal{C}_{e_i})|-4,2)\right\}+|E_i|$ starts to be greater than 2, $e_i$ will continue evolving to violate the prescribed performance bounds. Therefore, suppose that there exists $e_i\in \mathcal{E}^f$ satisfying $\sum\limits_{\mathcal{C}_{e_i}\in \mathcal{X}_i}\left\{\min (|\mathcal{E}(\mathcal{C}_{e_i})|-4,2)\right\}+|E_i|\geq 3$, then no matter  how we design the leaders in $\mathcal{G}$, in the worst case $e_i$ will always evolve to violate the prescribed performance bound, which leads to a contradiction. Hence, we can conclude that $\mathcal{G}$ should not contain a leaderless induced subgraph $\mathcal{G}^f=(\mathcal{V}^f,\mathcal{E}^f)$ such that there exists $e_i\in \mathcal{E}^f$ satisfying $\sum\limits_{\mathcal{C}_{e_i}\in \mathcal{X}_i}\left\{\min (|\mathcal{E}(\mathcal{C}_{e_i})|-4,2)\right\}+|E_i|\geq 3$. Or in other words, every leaderless graph $\mathcal{G}^f=(\mathcal{V}^f,\mathcal{E}^f)$ such that $\mathcal{G}^f \subseteq \mathcal{G} $, should satisfy \eqref{eq:ne_cycle}.\qed
\end{pf}
\begin{rem}
Lemma \ref{lem:1} and Theorem \ref{thm:1} indicate that when considering the leader-follower multi-agent system $\Sigma$ to achieve the target formation within the prescribed performance bounds, we should assign leaders such that \eqref{eq:ne_cycle} holds. They also propose a criterion in choosing leaders to achieve the target formation with prescribed performance guarantees, which can be further applied to solve leader selection problems. Note that Lemma \ref{lem:1} is a specific case of Theorem \ref{thm:1}. That is, when the graph is a tree, then the term $\sum\limits_{\mathcal{C}_{e_i}\in \mathcal{X}_i}\left\{\min (|\mathcal{E}(\mathcal{C}_{e_i})|-4,2)\right\}$ vanishes and $|E_i|$ is the number of edges that share a node with $e_i$, i.e., $|E_i|= |\mathcal{N}(e_i)|$. This also leads to the condition   $|\mathcal{N}(e_i)|\leq 2$ for tree graphs as in \eqref{eq:ne_tree}. 
Condition \eqref{eq:ne_tree} of Lemma \ref{lem:1} indicates that for a leaderless tree graph, each edge can have at most $2$ neighboring edges, which also means that each agent can have at most $3$ neighboring agents. Furthermore, according to Theorem \ref{thm:1}, for a general leaderless graph, agents that are belonging to some cycles can have more neighboring agents when including more cycles as long as condition \eqref{eq:ne_cycle} holds.  
\end{rem}

\section{Necessary and Sufficient Conditions on Graph Topology}
In the previous section, we derived necessary conditions on the graph topology that need to hold for both tree graphs and general graphs with cycles. We next derive sufficient conditions on the graph topology for both tree graphs and general graphs with cycles in this section. First of all, note that the conditions in Lemma \ref{lem:1} and Theorem \ref{thm:1} are necessary but not sufficient conditions since we only check the induced subgraphs (as in Definition 1) that are composed of followers. However, we have not taken the choice of leader vertices into account, which show the couplings between different induced subgraphs with only followers. In fact, for a general graph satisfying the necessary conditions in Lemma \ref{lem:1} or Theorem \ref{thm:1}, we still may not find a suitable PPC law for the leaders to steer the leader-follower multi-agent system to achieve the target formation within the prescribed performance bounds because the conditions in Lemma \ref{lem:1} or Theorem \ref{thm:1} are not sufficient when taking the leaders into account. In the sequel, we take the choice of leader vertices into account and focus on finding a sufficient condition on the graph topology, which will show to head to a necessary and sufficient condition. Consider the leader-follower multi-agent system under the communication graph $\mathcal{G}=(\mathcal{V},\mathcal{E})$, we first define the following follower-leader-follower (FLF) path and maximum follower-end subgraph as 
\begin{defn}\label{def:FLF}
(FLF path.) A path subgraph $p=(\mathcal{V}^p,\mathcal{E}^p)\in P$ of $\mathcal{G}=(\mathcal{V},\mathcal{E})$ as in Definition \ref{def:path} with $v_i,v_j\in \mathcal{V}^p$ as the two end nodes, is a FLF path of $\mathcal{G}$  if  $\mathcal{V}_\f^p=\{v_i,v_j\}$. $P$ is the collection of all the FLF paths of $\mathcal{G}$. The neighborhood of the FLF path $p\in P$ denoted as $\mathfrak{N}(p)$ is the set of edges that share the node $v_i$ or $v_j$ with $p$, i.e., $\mathfrak{N}(p)=\{e_k \mid v_i\in \mathbf{v}(e_k) \text{ or } v_j\in \mathbf{v}(e_k), e_k\in \mathcal{E}\setminus \mathcal{E}^p \}$. 
\end{defn}
\begin{defn}\label{def:max}
(Maximum follower-end subgraph.)
A graph $\mathcal{G}^{\star}=(\mathcal{V}^{\star},\mathcal{E}^{\star})$ is a maximum follower-end subgraph of the graph $\mathcal{G}=(\mathcal{V},\mathcal{E})$, denoted as $\mathcal{G}^{\star}\preceq \mathcal{G}$ if the following conditions hold:
\begin{itemize}
    \item $\mathcal{G}^{\star}\subseteq \mathcal{G}$ (as in Definition 1);
    \item every $i\in \mathcal{V}_\ld^{\star}$ belongs to a FLF path of $\mathcal{G}^{\star}$;
    \item there is no subgraph $\mathcal{G}^{\prime}=(\mathcal{V}^{\prime},\mathcal{E}^{\prime})$ of $\mathcal{G}$ such that every $i\in \mathcal{V}_\ld^{\prime}$ belongs to a FLF path satisfying $|\mathcal{V}^{\prime}|>|\mathcal{V}^{\star}|$.
\end{itemize}
\end{defn}
Definition \ref{def:max} is proposed in order to derive an induced subgraph $\mathcal{G}^{\star}$ of $\mathcal{G}$ (first requirement) such that the end leaders are ignored (second requirement) and includes as many agents as possible (third requirement). The insight here is that when a leader is placed in the graph as an end node, then this leader does not show any couplings between agents since it only connects to one agent and can freely move. For the maximum follower-end subgragh $\mathcal{G}^\star$, we can analyse the convergence of $e_i\in \mathcal{E}_{\f\f}^\star$ by resorting to the ideas and arguments from Lemma \ref{lem:1} and Theorem \ref{thm:1}. Note that we need to further consider the convergence of the dynamics of the leader-follower and leader-leader edge $e_i\in \mathcal{E}_{\ld\f}^\star\cup \mathcal{E}_{\ld\ld}^\star$, and the collection of all the FLF paths $P$ will traverse all the edges $e_i\in \mathcal{E}_{\ld\f}^\star\cup \mathcal{E}_{\ld\ld}^\star$, thus we next discuss the convergence results for FLF paths. When considering the maximum follower-end subgragh $\mathcal{G}^\star$, we can also completely decompose $\mathcal{G}^\star$ with respect to a specific FLF path $p_i\in P$. The complete decomposition of a graph $\mathcal{G}$ with respect to a specific FLF path $p_i$ is defined in a similar manner to Definition \ref{def:cd_edge}, and is given as follows:

\begin{defn}\label{def:cd_path}
(Complete decomposition w.r.t. FLF path.)
A graph $\mathcal{G}=(\mathcal{V},\mathcal{E})$ is decomposed with respect to the FLF path $p_i\in P$ as $\mathcal{G}:=\cup_{\mathcal{C}_{p_i}\in \mathcal{Y}_i}\mathcal{C}_{p_i}\cup \mathcal{Q}_i$, where $\mathcal{Y}_i:=\{\mathcal{C}_{p_i}\mid p_i\subseteq \mathcal{C}_{p_i}\}$ is the cycle set composed of all the cycles $\mathcal{C}_{p_i}$ in $\mathcal{G}$ that contain $p_i$ as per Definition \ref{def:path} and satisfy:
\begin{itemize}
\item for every pair of $\mathcal{C}_{p_i}^a, \mathcal{C}_{p_i}^b\in \mathcal{Y}_i$, $(\mathfrak{N}(p_i)\cap\mathcal{C}_{p_i}^a)\cap (\mathfrak{N}(p_i)\cap\mathcal{C}_{p_i}^b)=\emptyset$.
\item for every $\mathcal{C}_{p_i} \in \mathcal{Y}_i$, there does not exist a cycle $\mathcal{C}$ of $\mathcal{G}$ such that $p_i\subseteq \mathcal{C}, (\mathfrak{N}(p_i)\cap\mathcal{C}_{p_i})\cap (\mathfrak{N}(p_i)\cap\mathcal{C})\neq \emptyset$, and $|\mathcal{E}(\mathcal{C})|<|\mathcal{E}(\mathcal{C}_{p_i})|$,
\end{itemize}
and where $\mathcal{Q}_i:=\{e_k\mid e_k\notin \mathcal{C}_{p_i}, \mathcal{C}_{p_i} \in \mathcal{Y}_i\}$ is the set of the edges that do not belong to any cycle in $\mathcal{Y}_i$. Then, we call this decomposition a complete decomposition of $\mathcal{G}$ with respect to the FLF path $p_i\in P$.  \end{defn}
Now, we derive the following theorem regarding the necessary and sufficient conditions on the graph topology of the leader-follower multi-agent system in order to achieve the target formation $\mathcal{F}$ as in \eqref{eq:formation} under the control \eqref{eq:control} while satisfying the prescribed performance bounds \eqref{eq:pfp}.  
\begin{thm}\label{thm:2}
Consider the leader-follower multi-agent system $\Sigma$ described by the graph $\mathcal{G}=(\mathcal{V},\mathcal{E})$, and let $\mathcal{G}^\star \preceq \mathcal{G}$. A necessary and sufficient condition on $\mathcal{G}$ under which we can design the leaders using \eqref{eq:control} to achieve the target formation $\mathcal{F}$ as in \eqref{eq:formation} while satisfying \eqref{eq:pfp} is that:
\begin{itemize}
    \item for any $e_i\in \mathcal{E}^{\star}_{\f\f}$,
    \begin{equation}\label{eq:main1}
    \sum\limits_{\mathcal{C}_{e_i}\in \mathcal{X}_i}\left\{\min (|\mathcal{E}(\mathcal{C}_{e_i})|-4,2)\right\}+|E_i|\leq 2;
\end{equation}
\item for any $p_i\in P^{\star}$, there either exists $\mathcal{C}_{p_i}\in \mathcal{Y}_i$ such that $|\mathcal{E}(\mathcal{C}_{p_i})|<2|\mathcal{E}(p_i)|$, or
\begin{equation}\label{eq:main2}
  \sum\limits_{\mathcal{C}_{p_i}\in \mathcal{Y}_i}\left\{\min(|\mathcal{E}(\mathcal{C}_{p_i})|-2|\mathcal{E}(p_i)|-2,2)\right\}+|F_i|\leq 2,  
\end{equation}
\end{itemize}
where $E_i=\{e_k\mid e_k \in \mathcal{N}(e_i)\cap \mathcal{P}_i \}$, $\mathcal{X}_i$ is the cycle set obtained via the complete decomposition of $\mathcal{G}^\star$ with respect to $e_i\in \mathcal{E}^{\star}_{\f\f}$ as in Definition \ref{def:cd_edge}, 
$P^\star$ is the collection of all FLF paths of $\mathcal{G}^\star$, $ F_i=\{e_k\mid e_k \in \mathfrak{N}(p_i)\cap \mathcal{Q}_i \}$, and $\mathcal{Y}_i$ is the cycle set obtained via the complete decomposition of $\mathcal{G}^\star$ with respect to $p_i\in P^\star$ as in Definition \ref{def:cd_path}.
\end{thm}
\begin{pf}
We first prove the sufficient part. That is, assuming that the conditions \eqref{eq:main1} or \eqref{eq:main2} hold, we show that we can design the leaders using \eqref{eq:control} to achieve the target formation $\mathcal{F}$ as in \eqref{eq:formation} while satisfying \eqref{eq:pfp}. Firstly, if $\sum\limits_{\mathcal{C}_{e_i}\in \mathcal{X}_i}\left\{\min (|\mathcal{E}(\mathcal{C}_{e_i})|-4,2)\right\}+|E_i|\leq 2$ holds for every edge $e_i\in \mathcal{E}^{\star}_{\f\f}$, then in the worst case, when $\bar{x}_i$ is arbitrarily close to the prescribed performance bound, it will not evolve to violate the performance bound. Therefore, there always exists a positive, smooth and strictly decreasing performance function $\rho_{\bar{x}_i}(t)$ that can be predefined such that the dynamics of the edge $e_i\in \mathcal{E}^{\star}_{\f\f}$ will evolve while satisfying \eqref{eq:pfp}. Next, suppose that \eqref{eq:main2} holds, we first denote an arbitrary FLF path $p_i\in  P^{\star}$ by the consecutive sequence of nodes as $v_1,\cdots, v_{n_p}$ or the consecutive sequence of edges as $e_1,\cdots, e_{n_p-1}$ (a path with $n_p$ nodes and $n_p-1$ edges) with $v_1,v_{n_p}$ as the two followers and $e_1,e_{n_p-1}$ as the two leader-follower edges of path $p_i\in  P^{\star}$. If there exists a cycle $\mathcal{C}_{p_i}^1\in \mathcal{Y}_i$ that contains the FLF path $p_i$ as a subgraph and satisfies $|\mathcal{E}(\mathcal{C}_{p_i})|<2|\mathcal{E}(p_i)|$, then we know that the nodes $v_1,\cdots, v_{n_p}$ and the edges $e_1,\cdots, e_{n_p-1}$ of $p_i$ are also the respective nodes and edges of $\mathcal{C}_{p_i}^1$. Without loss of generality, suppose that the nodes of the cycle $\mathcal{C}_{p_i}^1$ are $v_1,\cdots, v_{n_p}, v_{n_p+1}, \cdots, v_{n_c}$ and the edges of the cycle $\mathcal{C}_{p_i}^1$ are $e_1,\cdots, e_{n_p-1}, e_{n_p}, \cdots, e_{n_c}$ (a cycle with $n_c$ nodes and $n_c$ edges), then we denote the remaining edges of $\mathcal{C}_{p_i}^1$ besides $e_1,\cdots, e_{n_p-1}$ as $e_{n_p},\cdots, e_{n_c}$. Since $e_1,\cdots, e_{n_p-1}$ and $e_{n_p},\cdots, e_{n_c}$ form the cycle $\mathcal{C}_{p_i}^1$, we have that $\sum_{i=1}^{n_p-1} \bar{x}_i+\sum_{j=n_p}^{n_c} \bar{x}_j=0$ or $\sum_{i=1}^{n_p-1} \bar{x}_i=-\sum_{j={n_p}}^{n_c} \bar{x}_j$. Since we have $|\mathcal{E}(\mathcal{C}_{p_i})|<2|\mathcal{E}(p_i)|$, i.e., $n_c<2(n_p-1)$, then the number of edges of $e_1\cdots e_{n_p-1}$ ($n_p-1$ edges) are more than that of the remaining edges $e_{n_p}\cdots e_{n_c}$ ($n_c-n_p+1$ edges). Therefore, $\sum_{i=1}^{n_p-1} \bar{x}_i$ will never evolve to violate the performance bound (the performance bound for $\sum_{i=1}^{n_p-1} \bar{x}_i$ is the summation of the performance bounds of each $\bar{x}_i$)  since it is constrained by $e_{n_p}\cdots e_{n_c}$ to form the cycle $\mathcal{C}_{p_i}^1$. Otherwise, if all $\mathcal{C}_{p_i}\in \mathcal{Y}_i$ satisfy $|\mathcal{E}(\mathcal{C}_{p_i})|\geq 2|\mathcal{E}(p_i)|$,
we can further calculate the dynamics of $p_i$ as $\sum_{i=1}^{n_p-1}\dot{\bar{x}}_i=-(\bar{x}_1+\bar{x}_{n_p-1})+\sum\nolimits_{\{j\mid e_j\in \mathfrak{N}(p_i)\}}\bar{x}_j$. Similar to the arguments in Theorem \ref{thm:1} when discussing the cycles, the first term of $-(\bar{x}_1+\bar{x}_{n_p-1})$ contributes the decay rate of $-2$ in the dynamics of $p_i$. The second term corresponds to the neighboring edges of $p_i$, which include the edges belonging to some cycles that contain $p_i$ and the remaining edges defined by $F_i$. We first consider the case when there exists only one cycle $\mathcal{C}_{p_i}^1$ that contains $p_i$ as a subgraph, $e_1\cdots e_{n_p-1}$ are the edges of $p_i$, $\mathfrak{N}(p_i)\cap \mathcal{C}_{p_i}^1=\{e_{n_p},e_{n_c}\}$ are the two neighboring edges of $p_i$ in the cycle $\mathcal{C}_{p_i}^1$, and the remaining edges are $e_{n_p+1},\cdots, e_{n_c-1}$. Then the dynamics of $p_i$ are rewritten as $\sum_{i=1}^{n_p-1}\dot{\bar{x}}_i=-(\bar{x}_1+\bar{x}_{n_p-1})+\bar{x}_{n_p}+\bar{x}_{n_c}+\sum\nolimits_{\{j\mid e_k\in F_i\}}\bar{x}_k$. We know that $\sum_{i=1}^{n_p-1}\bar{x}_i+\bar{x}_{n_p}+\bar{x}_{n_c}+\sum_{j=n_p+1}^{n_c-1}\bar{x}_j=0$ as they form the cycle $\mathcal{C}_{p_i}^1$, thus $\bar{x}_{n_p}+\bar{x}_{n_c}=-\sum_{i=1}^{n_p-1}\bar{x}_i-\sum_{j=n_p+1}^{n_c-1}\bar{x}_j$ and the dynamics of $p_i$ are  $\sum_{i=1}^{n_p-1}\dot{\bar{x}}_i=-(\bar{x}_1+\bar{x}_{n_p-1})-\sum_{i=1}^{n_p-1}\bar{x}_i-\sum_{j=n_p+1}^{n_c-1}\bar{x}_j+\sum\nolimits_{\{j\mid e_k\in F_i\}}\bar{x}_k$. We have that the number of edges of $e_1\cdots e_{n_p-1}$ is $n_1=|\mathcal{E}(p_i)|=n_p-1$ and the number of edges of $e_{n_p+1}\cdots e_{n_c-1}$ is $n_2=|\mathcal{E}(\mathcal{C}_{p_i}^1)|- |\mathcal{E}(p_i)|-2=n_c-n_p-1$. When $n_2-n_1\leq 2$, the term $-\sum_{i=1}^{n_p-1}\bar{x}_i-\sum_{j=n_p+1}^{n_c-1}\bar{x}_j$ contributes a decay rate of $n_2-n_1$ in the dynamics of $p_i$, while $n_2-n_1> 2$, the term $-\sum_{i=1}^{n_p-1}\bar{x}_i-\sum_{j=n_p+1}^{n_c-1}\bar{x}_j$ contributes a decay rate of $2$ in the dynamics of $p_i$ due to the constraints to form the cycle $\mathcal{C}_{p_i}^1$. In addition the term $\sum\nolimits_{\{j\mid e_k\in F_i\}}\bar{x}_k$ contributes a decay rate of $|F_i|$ in the dynamics of $p_i$. Hence, in the worst case the decay rate of the dynamics of $p_i$ is $\sum_{i=1}^{n_p-1}\dot{\bar{x}}_i=-2+\min (n_2-n_1,2)+|F_i|=-2+\min(|\mathcal{E}(\mathcal{C}_{p_i}^1)|-2|\mathcal{E}(p_i)|-2,2)+|F_i|\leq 0$ according to condition \eqref{eq:main2}. Next, we extend the analysis to the cases when there are more than one cycle that contain $p_i$ as a subgraph. Similarly then the decay rate of the dynamics of $p_i$ is $\sum_{i=1}^{n_p-1}\dot{\bar{x}}_i=-2+\sum\limits_{\mathcal{C}_{p_i}\in \mathcal{Y}_i}\left\{\min(|\mathcal{E}(\mathcal{C}_{p_i})|-2|\mathcal{E}(p_i)|-2,2)\right\}+|F_i|\leq 0$ according to condition \eqref{eq:main2}. This means that in the worst case, when condition \eqref{eq:main2} holds, we can always ensure that when each edge of $p_i\in P^\star$ is arbitrarily close to the prescribed performance bound, it will not evolve to violate the performance bound. Then, for the path $p_i$ which only has two followers as the end nodes and the remaining nodes are all leaders, we can apply the control law \eqref{eq:control} for the leaders in $p_i$ to ensure that every edge in $p_i$ satisfies the prescribed performance bound \eqref{eq:pfp}. This is generally  due to the fact that each edge of a path graph has at most 2 neighboring edges. We also refer the readers to \citep{chen2020leader} for the convergence analysis of the specific path graphs. Finally, since $P^\star$ is the collection of all FLF paths, it includes all the edges $e_i\in \mathcal{E}^{\star}_{\ld\ld}$ and all the edges $e_i\in \mathcal{E}^{\star}_{\ld\f}$. Note that since condition \eqref{eq:main1} is for all the edges $e_i\in \mathcal{E}^{\star}_{\f\f}$, while condition \eqref{eq:main2} considers all the edges $e_i\in \mathcal{E}^{\star}_{\ld\ld}\cup \mathcal{E}^{\star}_{\ld\f}$, these two conditions are independent of each other. Till now, we have considered all the edges $e_i\in \mathcal{E}^{\star}=\mathcal{E}^{\star}_{\f\f}\cup \mathcal{E}^{\star}_{\ld\f} \cup \mathcal{E}^{\star}_{\ld\ld}$ and the dynamics of all the edges satisfy the prescribed performance bound \eqref{eq:pfp}. Therefore, these two conditions are sufficient such that the leader-follower multi-agent system can achieve the target formation $\mathcal{F}$ as in \eqref{eq:formation} while satisfying the prescribed performance bound \eqref{eq:pfp}.

For the necessity part, we use contradiction. That is, suppose that either condition \eqref{eq:main1} or condition \eqref{eq:main2} does not hold. Firstly, suppose that there exists an edge $e_i\in \mathcal{E}^{\star}_{FF}$ such that $\sum\limits_{\mathcal{C}_{e_i}\in \mathcal{X}_i}\left\{\min (|\mathcal{E}(\mathcal{C}_{e_i})|-4,2)\right\}+|E_i|\geq 3$. Then, if $e_i$ and its neighboring edges are arbitrarily close to the prescribed performance bound, $e_i$ will continue to evolve to violate the performance bound. This results in a contradiction. Similarly, suppose that there exists a FLF path $p_i\in P^\star$ such that condition \eqref{eq:main2} does not hold. Then, if the edges of $p_i$ and the neighboring edges of $p_i$ are arbitrarily close to the prescribed performance bound, $p_i$ will continue to evolve to violate the performance bound, which also leads to a contradiction. Note that the performance bound for $p_i$ is the summation of the performance bounds of each edge of $p_i$. Finally, we can conclude that in order to achieve the target formation $\mathcal{F}$ as in \eqref{eq:formation} while satisfying \eqref{eq:pfp} by applying the control law \eqref{eq:control}, conditions \eqref{eq:main1} and \eqref{eq:main2} should hold. Therefore, conditions \eqref{eq:main1} and \eqref{eq:main2} are also necessary.\qed 
\end{pf}
\begin{rem}
Theorem \ref{thm:2} presents a necessary and sufficient condition on the leader-follower graph topology such that we can further design the leaders in order to steer the entire system to achieve the target formation within the transient constraints bounds. This also proposes a methodology for leader selection problems. Conditions \eqref{eq:main1} and \eqref{eq:main2} indicate the trade-offs among the leader-follower topology, number of cycles, number and positions of the leaders. Note that if we generalize the follower-follower edge also as a FLF path, i.e., a path that starts with a follower, ends with a follower, but crosses no leaders, then, the first condition in Theorem \ref{thm:2} can be regarded as a special case of the second condition, since $|\mathcal{E}(\mathcal{C}_{e_i})|<2|\mathcal{E}(e_i)|=2$ will never hold for $e_i\in \mathcal{E}^{\star}_{FF}$, and $\min(|\mathcal{E}(\mathcal{C}_{e_i})|-2|\mathcal{E}(e_i)|-2,2)$ is exactly $\min (|\mathcal{E}(\mathcal{C}_{e_i})|-4,2)$. Similar to the discussions on Lemma \ref{lem:1} and Theorem \ref{thm:1} in the previous section, Theorem \ref{thm:2} holds for both tree graphs and general graphs with cycles. 
\end{rem}

\section{Simulations and Examples}
\begin{figure}[!h]
\centering
\includegraphics[width=0.85\columnwidth]{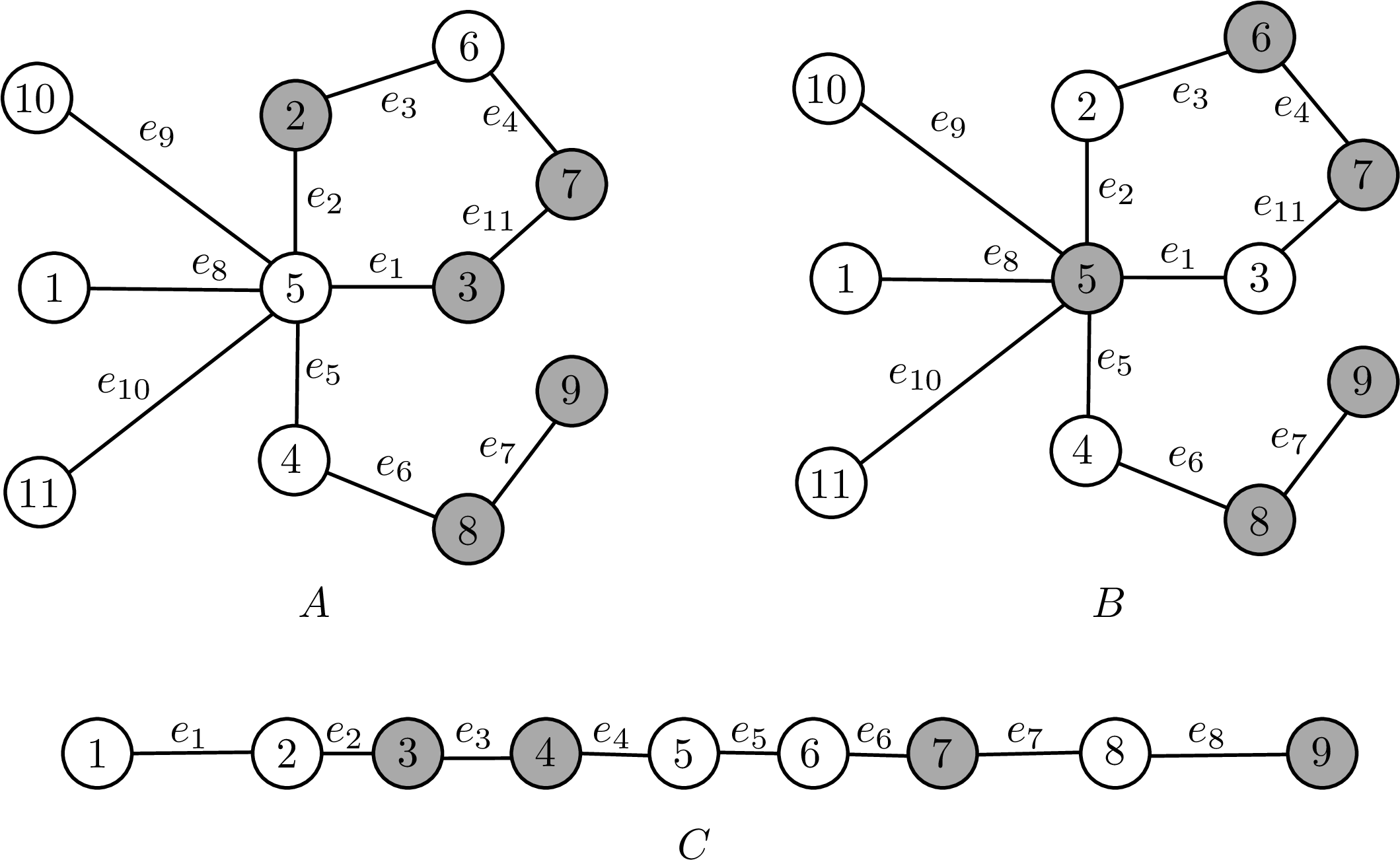}
\caption{Leader-follower communication graphs.}
\label{communication}
\end{figure}

In this section, simulation examples are presented in order to verify the
results. We first consider a multi-vehicle platooning example, and later on a multi-robot coordination example with different choice of leader robots. The simulations' communication graphs are shown in Fig. \ref{communication}, where the leaders and followers are represented by grey and white nodes, respectively. 

\begin{figure}[!h]
\centering
\includegraphics[width=0.9\columnwidth]{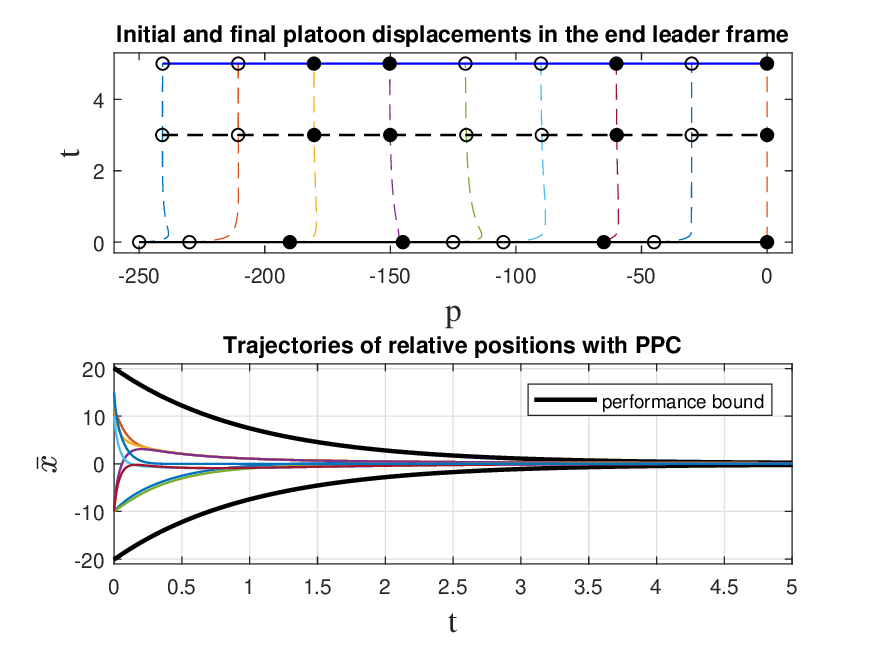}
\caption{Platoon displacements (top) and trajectories of relative positions (bottom).}
\label{platoon}
\end{figure}

\textbf{Multi-vehicle platooning:} we consider a vehicle platoon with $9$ vehicles as shown in Fig. \ref{communication}.$C$ with leader vehicles $\mathcal{V}_\ld=\{v_3,v_4,v_7,v_9\}.$ The vehicles are initialized with $p=[0, 20, 60, 105, 125, 145, 185, 205, 250]^T$.
All edges are subject to the transient constraints as in \eqref{eq:pfp} with $\rho_{\bar{x}_i}(t)=20e^{-t}+0.1$ and $\bar{p}^{des}_i=30$. The transient constraints under consideration encapsulate two primary tasks: ensuring collision avoidance with a minimum distance of 10 and maintaining connectivity within a maximum distance of 50. Furthermore, these constraints prescribe an additional requirement on the transient behavior, particularly concerning the convergence rate and overshoot towards the target platoon formation. Therefore, the vehicle platoon aims to achieve the formation (keeping a safe distance of $30$ between neighboring vehicles) while satisfying the time-varying constraints by only designing the controllers of the leader vehicles.  We can check that the necessary and sufficient conditions proposed in Theorem \ref{thm:2} are trivially satisfied for multi-vehicle platooning since each vehicle has at most $2$ neighboring vehicles. The simulation results when applying the PPC strategy \eqref{eq:control}  are depicted in Fig. \ref{platoon}. The top figure plots the platoon evolutions (dashed curve) within the end leader (indexed by $9$) frame. The initial platoon is shown in black while the final platoon is shown in blue which achieves a distance of $30$ between neighboring vehicles. We can verify that all the trajectories of the relative positions of the neighboring vehicles evolve within the transient performance bounds as shown in the bottom figure. Therefore, we can conclude that the target multi-vehicle platoon is achieved while satisfying the transient constraints. 

\begin{figure}
\begin{subfigure}{\columnwidth}
\centering
\includegraphics[width=\columnwidth,trim={2cm 0cm 2.5cm 0cm},clip]{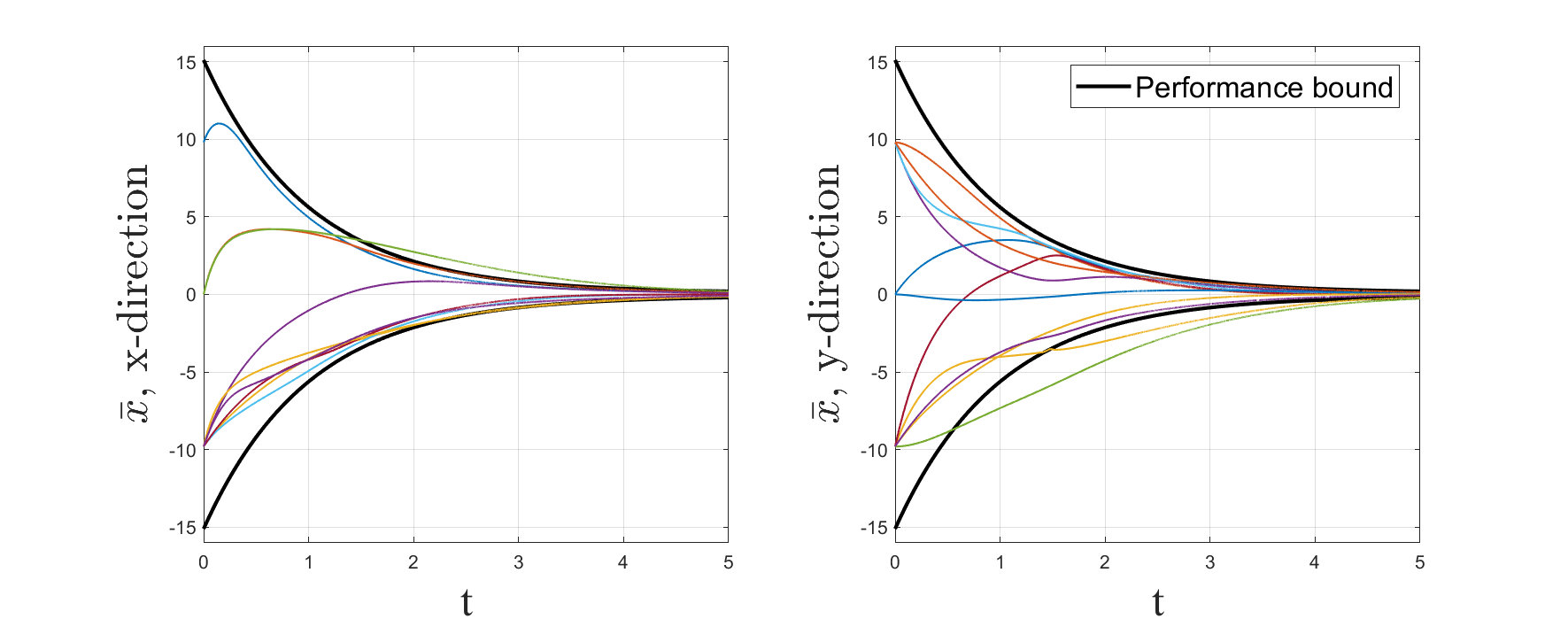}
\caption{Trajectories of the shifted relative positions under graph as in Fig. \ref{communication}.$A$.}
\label{exmp_3}
\end{subfigure}
\newline
\begin{subfigure}{\columnwidth}
\centering
\includegraphics[width=\columnwidth,trim={2cm 0cm 2.5cm 0cm},clip]{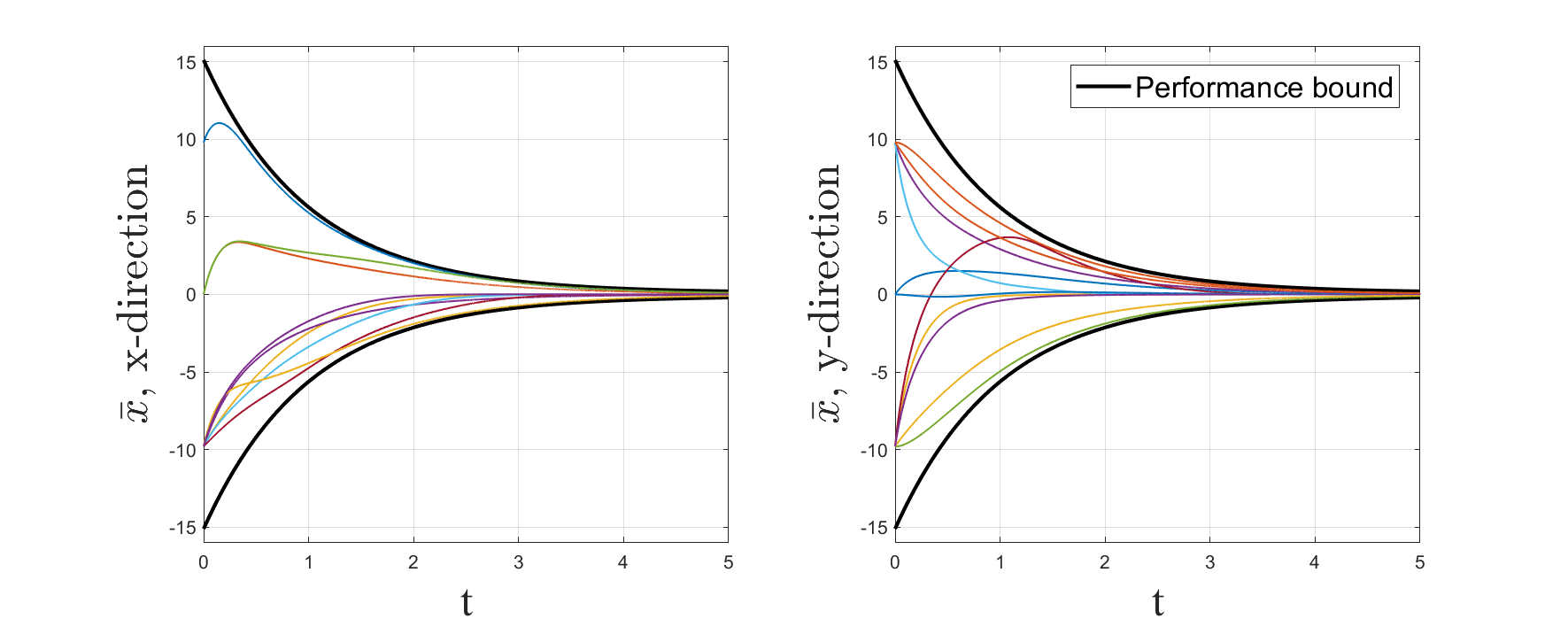}
\caption{Trajectories of the shifted relative positions under graph as in Fig. \ref{communication}.$B$.}
\label{exmp_4}
\end{subfigure}
\caption{Trajectories of the shifted relative positions.  }
\label{exmp_34}
\end{figure}

\textbf{Multi-robot coordination:} we then consider a multi-robot coordination example in $2$ dimensions under more complex communication graphs, and we show that different choices of leader robots influence the fulfillment of the transient constraints. Suppose all robots are initialized at the origin and the target relative position-based formation is $p^{des}_{53}=[-9.8,0]^T, p^{des}_{52}=[0,-9.8]^T, p^{des}_{62}=[9.8,9.8]^T, p^{des}_{76}=[9.8,-9.8]^T, p^{des}_{54}=[0,9.8]^T, p^{des}_{84}=[9.8,-9.8]^T, p^{des}_{98}=[9.8,9.8]^T, p^{des}_{51}=[9.8,0]^T,p^{des}_{510}:=p^{des}_{5}-p^{des}_{10}=[9.8,-9.8]^T,p^{des}_{511}:=p^{des}_{5}-p^{des}_{11}=[9.8,9.8]^T$. We choose, without loss of generality, the same $\rho_{\bar{x}_i}$ for all edges as $\rho_{\bar{x}_i}(t)=15e^{-t}+0.1$ and the same transformed function $T_{\bar{x}_i}$ as $ T_{\bar{x}_i}(\hat{\bar{x}}_{i})=\ln \left(\frac{1+\hat{\bar{x}}_{i}}{1-\hat{\bar{x}}_{i}}\right)$. Note that the presented example generally covers a multi-robot dispersion task and the transient constraints encompass both the maintenance of connectivity and the transient performance requirements. Next, we focus on verifying the results derived in Theorem \ref{thm:2}, and we consider different positions of the leader robots, which are depicted in Fig. \ref{communication}.$A$ and Fig. \ref{communication}.$B$ respectively. In Fig. \ref{communication}.$A$, the graph is described by $\mathcal{G}=(\mathcal{V},\mathcal{E})$ with $\mathcal{V}_\f=\{v_1,v_4,v_5,v_6,v_{10},v_{11}\}$ and $\mathcal{V}_\ld=\{v_2,v_3,v_7,v_8,v_9\}$. The edge set is $\mathcal{E}=\{e_1,\dots, e_{11}\}$. We first construct the maximum follower-end subgraph  $\mathcal{G}^{\star}=(\mathcal{V}^{\star},\mathcal{E}^{\star})$ of $\mathcal{G}$ according to Definition \ref{def:max}, and obtain that $\mathcal{V}^{\star}=\mathcal{V}\setminus \{v_8,v_9\}$, $\mathcal{E}^{\star}=\mathcal{E}\setminus \{e_6,e_7\}$. With respect to $\mathcal{G}^{\star}$, we have $\mathcal{E}^{\star}_{\f\f}=\{e_5,e_8,e_9,e_{10}\}$. For these follower-follower edges, we obtain $\sum\limits_{\mathcal{C}_{e_i}\in \mathcal{X}_i}\left\{\min (|\mathcal{E}(\mathcal{C}_{e_i})|-4,2)\right\}+|E_i|=|E_i|=5> 2, i\in \{5,8,9,10\}$, which means that condition \eqref{eq:main1} as in Theorem \ref{thm:2} does not hold. Furthermore, there exist $2$ FLF paths in $\mathcal{G}^{\star}$, i.e., $p_1=v_5v_2v_6$ and $p_2=v_5v_3v_7v_6$. For both FLF paths, we then operate complete decomposition according to Definition \ref{def:cd_path}. We can verify that condition \eqref{eq:main2} holds for $p_2$ since $|\mathcal{E}(\mathcal{C}_{p_2})|=5<2|\mathcal{E}(p_2)|=6$. However, for the FLF path $p_1$, we have that $\sum\limits_{\mathcal{C}_{p_1}\in \mathcal{Y}_1}\left\{\min(|\mathcal{E}(\mathcal{C}_{p_1})|-2|\mathcal{E}(p_1)|-2,2)\right\}+|F_1|=\min(5-4-2,2)+4=3> 2$ since $F_1=\{e_5,e_8,e_9,e_{10}\}$. Then, the simulation results when applying the PPC strategy \eqref{eq:control} for the leader-follower topology as in Fig. \ref{communication}.$A$ are shown in Fig.  \ref{exmp_3}. We can see that the performance bounds are violated since the necessary and sufficient conditions \eqref{eq:main1}, \eqref{eq:main2} in Theorem \ref{thm:2} do not hold. As a comparison, we change the positions of the leader robots as shown in Fig. \ref{communication}.$B$, where the new graph $\mathcal{G}=(\mathcal{V},\mathcal{E})$  has the respective follower and leader sets as $\mathcal{V}_\f=\{v_1,v_2,v_3,v_4,v_{10},v_{11}\}$ and $\mathcal{V}_\ld=\{v_5,v_6,v_7,v_8,v_9\}$. The edge set is $\mathcal{E}=\{e_1,\dots, e_{11}\}$. Similarly, we first construct the maximum follower-end subgraph  $\mathcal{G}^{\star}=(\mathcal{V}^{\star},\mathcal{E}^{\star})$ of $\mathcal{G}$ according to Definition \ref{def:max}, and derive that $\mathcal{V}^{\star}=\mathcal{V}\setminus \{v_8,v_9\}$, $\mathcal{E}^{\star}=\mathcal{E}\setminus \{e_6,e_7\}$. With respect to $\mathcal{G}^{\star}$, we now have $\mathcal{E}^{\star}_{\f\f}=\emptyset$ since there is no follower-follower edge. Then, we check whether condition \eqref{eq:main2} holds for all FLF paths. For the FLF paths $v_iv_5v_j$ in $\mathcal{G}^{\star}$ such that $i,j\in \{1,4,10,11\}$ and $i\neq j$, note that they do not have neighboring edges in $\mathcal{G}^{\star}$. For example, the FLF path $p_1=v_1v_5v_{11}$ does not have any neighboring edges in $\mathcal{G}^{\star}$ according to Definition \ref{def:FLF}. We can conclude that condition \eqref{eq:main2} holds for $p_1$ since $  \sum\limits_{\mathcal{C}_{p_1}\in \mathcal{Y}_1}\left\{\min(|\mathcal{E}(\mathcal{C}_{p_1})|-2|\mathcal{E}(p_1)|-2,2)\right\}+|F_1|=|F_1|=0< 2$. Next, for the FLF paths $v_iv_5v_j$ in $\mathcal{G}^{\star}$ such that $i\in \{1,4,10,11\}$ and $j\in \{2,3\}$, note that they have a single neighboring edge $e_3$ if $j=2$ or $e_{11}$ if $j=3$ in $\mathcal{G}^{\star}$. For example, the FLF path $p_2=v_1v_5v_{2}$ has only one neighboring edge in $\mathcal{G}^{\star}$, i.e., $e_3$. We can conclude that condition \eqref{eq:main2} holds for $p_2$ since $  \sum\limits_{\mathcal{C}_{p_2}\in \mathcal{Y}_2}\left\{\min(|\mathcal{E}(\mathcal{C}_{p_2})|-2|\mathcal{E}(p_2)|-2,2)\right\}+|F_2|=|F_2|=1< 2$. The remaining FLF paths are $p_3=v_2v_6v_7v_3$ and $p_4=v_2v_5v_3$. They both belong to the cycle composed by the edges $e_1,e_2,e_3,e_4,e_{11}$. We can verify that condition \eqref{eq:main2} holds for $p_3$ since $|\mathcal{E}(\mathcal{C}_{p_3})|=5<2|\mathcal{E}(p_3)|=6$. Moreover, for the FLF path $p_4$, we have that $\sum\limits_{\mathcal{C}_{p_4}\in \mathcal{Y}_4}\left\{\min(|\mathcal{E}(\mathcal{C}_{p_4})|-2|\mathcal{E}(p_4)|-2,2)\right\}+|F_4|=\min(5-4-2,2)=-1< 2$ since $F_4=\emptyset$. Then, the simulation results when applying the PPC strategy \eqref{eq:control} for the leader-follower topology as in Fig. \ref{communication}.$B$ are shown in Fig.  \ref{exmp_4}. We can observe that all the trajectories evolve within the prescribed performance bounds since the necessary and sufficient conditions \eqref{eq:main1}, \eqref{eq:main2} in Theorem \ref{thm:2} hold. We can conclude that for the leader-follower network as shown in Fig. \ref{communication}.$B$, since the necessary and sufficient conditions \eqref{eq:main1}, \eqref{eq:main2} in Theorem \ref{thm:2} hold, the target formation is achieved within the prescribed performance bounds (as shown in Fig. \ref{exmp_4}) when applying the PPC strategy \eqref{eq:control}.

\section{Conclusions}
In this paper, we have investigated conditions on the graph topology such that relative position-based formation  within prescribed transient bounds can be achieved for leader-follower multi-agent systems. For both tree graphs and general graphs with cycles, necessary conditions on the graph topology are firstly proposed under which we can potentially design the leaders to achieve the target formation with prescribed performance guarantees, which are further extended to necessary and sufficient conditions on the leader-follower graph topology such that the target formation can be achieved within the prescribed performance bounds. Future research includes considering other transient approaches and also investigating leader selection problems.


\bibliographystyle{agsm} 
\bibliography{autosam}           



\end{document}